


\font\mayusc=cmcsc10 

\def\jnp#1#2#3{\frenchspacing {\sl Nucl. Phys.}, {\bf B#1} ({#2}) {#3}}
\def\jpl#1#2#3{\frenchspacing {\sl Phys. Lett.}, {\bf B#1} ({#2}) {#3}}
\def\jpr#1#2#3{\frenchspacing {\sl Phys. Rev.}, {\bf #1} ({#2}) {#3}}
\def\jprl#1#2#3{\frenchspacing {\sl Phys. Rev. Lett.}, {\bf #1} ({#2}) {#3}}
\def\jzp#1#2#3{\frenchspacing {\sl Z. Phys.}, {\bf #1} ({#2}) {#3}}


      \font \ninebf                 = cmbx9
      \font \ninei                  = cmmi9
      \font \nineit                 = cmti9
      \font \ninerm                 = cmr9
      \font \ninesans               = cmss10 at 9pt
      \font \ninesl                 = cmsl9
      \font \ninesy                 = cmsy9
      \font \ninett                 = cmtt9
      \font \fivesans               = cmss10 at 5pt
						\font \sevensans              = cmss10 at 7pt
      \font \sixbf                  = cmbx6
      \font \sixi                   = cmmi6
      \font \sixrm                  = cmr6
						\font \sixsans                = cmss10 at 6pt
      \font \sixsy                  = cmsy6
      \font \tams                   = cmmib10
      \font \tamss                  = cmmib10 scaled 700
						\font \tensans                = cmss10
    
      \skewchar\ninei='177 \skewchar\sixi='177
      \skewchar\ninesy='60 \skewchar\sixsy='60
      \hyphenchar\ninett=-1
      \def\newline{\hfil\break}%
      \catcode`@=11
      \def\folio{\ifnum\pageno<\z@
      \uppercase\expandafter{\romannumeral-\pageno}%
      \else\number\pageno \fi}
      \catcode`@=12 

      \newfam\sansfam
      \textfont\sansfam=\tensans\scriptfont\sansfam=\sevensans
      \scriptscriptfont\sansfam=\fivesans
      \def\sans{\fam\sansfam\tensans}


      \def\petit{\def\rm{\fam0\ninerm}%
      \textfont0=\ninerm \scriptfont0=
\sixrm \scriptscriptfont0=\fiverm
       \textfont1=\ninei \scriptfont1=
\sixi \scriptscriptfont1=\fivei
       \textfont2=\ninesy \scriptfont2=
\sixsy \scriptscriptfont2=\fivesy
       \def\it{\fam\itfam\nineit}%
       \textfont\itfam=\nineit
       \def\sl{\fam\slfam\ninesl}%
       \textfont\slfam=\ninesl
       \def\bf{\fam\bffam\ninebf}%
       \textfont\bffam=\ninebf \scriptfont\bffam=\sixbf
       \scriptscriptfont\bffam=\fivebf
       \def\sans{\fam\sansfam\ninesans}%
       \textfont\sansfam=\ninesans \scriptfont\sansfam=\sixsans
       \scriptscriptfont\sansfam=\fivesans
       \def\tt{\fam\ttfam\ninett}%
       \textfont\ttfam=\ninett
       \normalbaselineskip=11pt
       \setbox\strutbox=\hbox{\vrule height7pt depth2pt width0pt}%
       \normalbaselines\rm


      \def\bvec##1{{\textfont1=\tbms\scriptfont1=\tbmss
      \textfont0=\ninebf\scriptfont0=\sixbf
      \mathchoice{\hbox{$\displaystyle##1$}}{\hbox{$\textstyle##1$}}
      {\hbox{$\scriptstyle##1$}}{\hbox{$\scriptscriptstyle##1$}}}}}


\def\imag{\mathop{\rm Im}}

.

					\mathchardef\gammav="0100
     \mathchardef\deltav="0101
     \mathchardef\thetav="0102
     \mathchardef\lambdav="0103
     \mathchardef\xiv="0104
     \mathchardef\piv="0105
     \mathchardef\sigmav="0106
     \mathchardef\upsilonv="0107
     \mathchardef\phiv="0108
     \mathchardef\psiv="0109
     \mathchardef\omegav="010A


					\mathchardef\Gammav="0100
     \mathchardef\Deltav="0101
     \mathchardef\Thetav="0102
     \mathchardef\Lambdav="0103
     \mathchardef\Xiv="0104
     \mathchardef\Piv="0105
     \mathchardef\Sigmav="0106
     \mathchardef\Upsilonv="0107
     \mathchardef\Phiv="0108
     \mathchardef\Psiv="0109
     \mathchardef\Omegav="010A



\font\grbfivefm=cmbx5
\font\grbsevenfm=cmbx7
\font\grbtenfm=cmbx10 
\newfam\grbfam
\textfont\grbfam=\grbtenfm
\scriptfont\grbfam=\grbsevenfm
\scriptscriptfont\grbfam=\grbfivefm

\font\calbfivefm=cmbsy10 at 5pt
\font\calbsevenfm=cmbsy10 at 7pt
\font\calbtenfm=cmbsy10 
\newfam\calbfam
\textfont\calbfam=\calbtenfm
\scriptfont\calbfam=\calbsevenfm
\scriptscriptfont\calbfam=\calbfivefm



      \def\bvec#1{{\textfont1=\tams\scriptfont1=\tamss
      \textfont0=\tenbf\scriptfont0=\sevenbf
      \mathchoice{\hbox{$\displaystyle#1$}}{\hbox{$\textstyle#1$}}
      {\hbox{$\scriptstyle#1$}}{\hbox{$\scriptscriptstyle#1$}}}}



\def\pmbf#1{\leavevmode\setbox0=\hbox{#1}%
\kern-.02em\copy0\kern-\wd0
\kern.04em\copy0\kern-\wd0
\kern-.02em\copy0\kern-\wd0
\kern-.03em\copy0\kern-\wd0
\kern.06em\box0 }



						\def\monthname{%
   			\ifcase\month
      \or Jan\or Feb\or Mar\or Apr\or May\or Jun%
      \or Jul\or Aug\or Sep\or Oct\or Nov\or Dec%
   			\fi
							}%
					\def\timestring{\begingroup
   		\count0 = \time
   		\divide\count0 by 60
   		\count2 = \count0   
   		\count4 = \time
   		\multiply\count0 by 60
   		\advance\count4 by -\count0   
   		\ifnum\count4<10
     \toks1 = {0}%
   		\else
     \toks1 = {}%
   		\fi
   		\ifnum\count2<12
      \toks0 = {a.m.}%
   		\else
      \toks0 = {p.m.}%
      \advance\count2 by -12
   		\fi
   		\ifnum\count2=0
      \count2 = 12
   		\fi
   		\number\count2:\the\toks1 \number\count4 \thinspace \the\toks0
					\endgroup}%

				\def\timestamp{\number\day\space
\monthname\space\number\year\quad\timestring}%
				\newskip\abovelistskip      \abovelistskip = .5\baselineskip 
				\newskip\interitemskip      \interitemskip = 0pt
				\newskip\belowlistskip      \belowlistskip = .5\baselineskip
				\newdimen\listleftindent    \listleftindent = 0pt
				\newdimen\listrightindent   \listrightindent = 0pt

				%


\def\petit{\def\rm{\fam0\ninerm}%
\textfont0=\ninerm \scriptfont0=\sixrm \scriptscriptfont0=\fiverm
\textfont1=\ninei \scriptfont1=\sixi \scriptscriptfont1=\fivei
\textfont2=\ninesy \scriptfont2=\sixsy \scriptscriptfont2=\fivesy
       \def\it{\fam\itfam\nineit}%
       \textfont\itfam=\nineit
       \def\sl{\fam\slfam\ninesl}%
       \textfont\slfam=\ninesl
       \def\bf{\fam\bffam\ninebf}%
       \textfont\bffam=\ninebf \scriptfont\bffam=\sixbf
       \scriptscriptfont\bffam=\fivebf
       \def\sans{\fam\sansfam\ninesans}%
       \textfont\sansfam=\ninesans \scriptfont\sansfam=\sixsans
       \scriptscriptfont\sansfam=\fivesans
       \def\tt{\fam\ttfam\ninett}%
       \textfont\ttfam=\ninett
       \normalbaselineskip=11pt
       \setbox\strutbox=\hbox{\vrule height7pt depth2pt width0pt}%
       \normalbaselines\rm
      \def\vec##1{{\textfont1=\tbms\scriptfont1=\tbmss
      \textfont0=\ninebf\scriptfont0=\sixbf
      \mathchoice{\hbox{$\displaystyle##1$}}{\hbox{$\textstyle##1$}}
      {\hbox{$\scriptstyle##1$}}{\hbox{$\scriptscriptstyle##1$}}}}}

      \def\footnoterule{\kern-3pt\hrule width 2cm\kern2.6pt}
      \newdimen\oldparindent\oldparindent=1.5em
      \parindent=1.5em
 
\newcount\footcount \footcount=0
\def\advftncnt{\advance\footcount by1\global\footcount=\footcount}
      \def\fnote#1{\advftncnt$^{\the\footcount}$\begingroup\petit
      \parfillskip=0pt plus 1fil
      \def\textindent##1{\hangindent0.5\oldparindent\noindent\hbox
      to0.5\oldparindent{##1\hss}\ignorespaces}%
 \vfootnote{$^{\the\footcount}$}
{#1\nullbox{0mm}{2mm}{0mm}\vskip-9.69pt}\endgroup}


      \def\item#1{\par\noindent
      \hangindent6.5 mm\hangafter=0
      \llap{#1\enspace}\ignorespaces}
      
      \def\leaderfill{\kern0.5em\leaders
\hbox to 0.5em{\hss.\hss}\hfill\kern
      0.5em}
						\def\hb{\hfill\break}

    \def\centerrule#1{\centerline{\kern#1\hrulefill\kern#1}}


      \def\boxit#1{\vbox{\hrule\hbox{\vrule\kern3pt
						\vbox{\kern3pt#1\kern3pt}\kern3pt\vrule}\hrule}}

      \def\tightboxit#1{\vbox{\hrule\hbox{\vrule
						\vbox{#1}\vrule}\hrule}}

      \def\looseboxit#1{\vbox{\hrule\hbox{\vrule\kern5pt
						\vbox{\kern5pt#1\kern5pt}\kern5pt\vrule}\hrule}}

      \def\youboxit#1#2{\vbox{\hrule\hbox{\vrule\kern#2
						\vbox{\kern#2#1\kern#2}\kern#2\vrule}\hrule}}




			\def\whitetile#1#2#3{\setbox0=\null
			\ht0=#1 \dp0=#2\wd0=#3 \setbox1=
\hbox{\tightboxit{\box0}}\lower#2\box1}

			\def\nullbox#1#2#3{\setbox0=\null
			\ht0=#1 \dp0=#2\wd0=#3\box0}




\def\fig{\leavevmode Fig.}

\def\equ{\leavevmode Eq.}

\def\sect{\leavevmode Sect.}
\def\subsect{\leavevmode Subsect.}

\def\equn#1{\ifmmode \eqno{\rm #1}\else \equ~#1\fi}



\def\tev{\ifmmode \mathop{\rm TeV}\nolimits\else {\rm TeV}\fi}
\def\gev{\ifmmode \mathop{\rm GeV}\nolimits\else {\rm GeV}\fi}
\def\mev{\ifmmode \mathop{\rm MeV}\nolimits\else {\rm MeV}\fi}
\def\kev{\ifmmode \mathop{\rm keV}\nolimits\else {\rm keV}\fi}
\def\ev{\ifmmode \mathop{\rm eV}\nolimits\else {\rm eV}\fi}

\def\chidof{\ifmmode
\mathop\chi^2/{\rm d.o.f.}\else $\chi^2/{\rm d.o.f.}\null$\fi}

\def\msbar{\ifmmode
\mathop{\overline{\rm MS}}\else$\overline{\rm MS}$\null\fi}


\def\physmatex{P\kern-.14em\lower.5ex\hbox{\sevenrm H}ys
\kern -.35em \raise .6ex \hbox{{\sevenrm M}a}\kern -.15em
 T\kern-.1667em\lower.5ex\hbox{E}\kern-.125emX\null}%

\def\ref#1{$^{[#1]}$\relax}

\def\prajnyp#1#2#3#4#5{
\frenchspacing{\mayusc #1}, {\sl#2}, {\bf #3}, {#5} {(#4)}}




\def\eqsub{\mathop{=}\limits}









\def\ddal{\mathop{\vrule height 7pt depth0.2pt
\hbox{\vrule height 0.5pt depth0.2pt width 6.2pt}
\vrule height 7pt depth0.2pt width0.8pt
\kern-7.4pt\hbox{\vrule height 7pt depth-6.7pt width 7.pt}}}
\def\sdal{\mathop{\kern0.1pt\vrule height 4.9pt depth0.15pt
\hbox{\vrule height 0.3pt depth0.15pt width 4.6pt}
\vrule height 4.9pt depth0.15pt width0.7pt
\kern-5.7pt\hbox{\vrule height 4.9pt depth-4.7pt width 5.3pt}}}
\def\ssdal{\mathop{\kern0.1pt\vrule height 3.8pt depth0.1pt width0.2pt
\hbox{\vrule height 0.3pt depth0.1pt width 3.6pt}
\vrule height 3.8pt depth0.1pt width0.5pt
\kern-4.4pt\hbox{\vrule height 4pt depth-3.9pt width 4.2pt}}}




\mathchardef\lap='0001


\def\lsim{\mathop{\setbox0=\hbox{$\displaystyle 
\raise2.2pt\hbox{$\;<$}\kern-7.7pt\lower2.6pt\hbox{$\sim$}\;$}
\box0}}
\def\gsim{\mathop{\setbox0=\hbox{$\displaystyle 
\raise2.2pt\hbox{$\;>$}\kern-7.7pt\lower2.6pt\hbox{$\sim$}\;$}
\box0}}

\def\gsimsub#1{\mathord{\vtop to0pt{\ialign{##\crcr
$\hfil{{\mathop{\setbox0=\hbox{$\displaystyle 
\raise2.2pt\hbox{$\;>$}\kern-7.7pt\lower2.6pt\hbox{$\sim$}\;$}
\box0}}}\hfil$\crcr\noalign{\kern1.5pt\nointerlineskip}
$\hfil\scriptstyle{#1}{}\kern1.5pt\hfil$\crcr}\vss}}}

\def\lsimsub#1{\mathord{\vtop to0pt{\ialign{##\crcr
$\hfil\displaystyle{\mathop{\setbox0=\hbox{$\displaystyle 
\raise2.2pt\hbox{$\;<$}\kern-7.7pt\lower2.6pt\hbox{$\sim$}\;$}
\box0}}
\def\gsim{\mathop{\setbox0=\hbox{$\displaystyle 
\raise2.2pt\hbox{$\;>$}\kern-7.7pt\lower2.6pt\hbox{$\sim$}\;$}
\box0}}\hfil$\crcr\noalign{\kern1.5pt\nointerlineskip}
$\hfil\scriptstyle{#1}{}\kern1.5pt\hfil$\crcr}\vss}}}

\def\ii{{\rm i}}
\def\dd{{\rm d}}

\def\gammae{\gamma_{\rm E}}




\def\frac#1#2{{#1\over#2}}
\def\dfrac#1#2{{\displaystyle{#1\over#2}}}
\def\tfrac#1#2{{\textstyle{#1\over#2}}}
\def\ffrac#1#2{\leavevmode
   \kern.1em \raise .5ex \hbox{\the\scriptfont0 #1}%
   \kern-.1em $/$%
   \kern-.15em \lower .25ex \hbox{\the\scriptfont0 #2}%
}%



\def\brochureb#1#2#3{\pageno#3
\headline={\ifodd\pageno{\rheadline}
\else\lheadline\fi}
\def\rheadline{\hfil -{#2}-\hfil}
\def\lheadline{\hfil-{#1}-\hfil}
\footline={\hss -- \number\pageno\ --\hss}
\voffset=2\baselineskip}

\def\nada{\phantom{M}\kern-1em}
\def\brochureendcover#1{\vfill\eject\pageno=1{\nada#1}\vfill\eject}





\def\chapterb#1#2#3{\pageno#3
\headline={\ifodd\pageno{\ifnum\pageno=#3\hfil\else\rheadline\fi}
\else\lheadline\fi}
\def\rheadline{\hfil -{#2}-\hfil}
\def\lheadline{\hfil-{#1}-\hfil}
\footline={\hss -- \number\pageno\ --\hss}
\voffset=2\baselineskip}


\def\bookendchapter{\ifodd\pageno
 \vfill\eject\footline={\hfill}\headline={\hfill}\null \vfill\eject
 \else\vfill\eject \fi}

\def\obookendchapter{\ifodd\pageno\vfill\eject
 \else\vfill\eject\null \vfill\eject\fi}


\def\booksection#1{
\setbox0=\vbox{\hsize=0.85\hsize\tolerance=500\raggedright\hfuzz=6mm
\noindent{\medfib #1}\medskip}\goodbreak\vskip0.6cm\box0
\nobreak
\noindent}
\def\booksubsection#1{
\setbox0=\vbox{\hsize=0.85\hsize\tolerance=400\raggedright\hfuzz=4mm
\noindent{\fib #1}\smallskip}\goodbreak\vskip0.45cm\box0
\nobreak
\noindent}




\def\figuresc#1#2{\petit{\noindent\sc#1}\ #2}

\def\captiontype{\tolerance=800\hfuzz=1mm\raggedright\noindent}



\def\abstracttype#1{
\hsize0.7\hsize\tolerance=800\hfuzz=0.5mm \noindent{\fib #1}\par
\medskip\petit}


\def\hb{\hfill\break}


\font\twelverm=cmr12 
\font\smallsc=cmcsc10 at 9pt 
\font\fib=cmfib8
\font\medfib=cmfib8 at 9pt


\font\sc=cmcsc10 

\font\addressfont=cmbxti10 at 9pt


\catcode`@=11 

\newdimen\pagewidth \newdimen\pageheight \newdimen\ruleht
 \maxdepth=2.2pt  \parindent=3pc
\pagewidth=\hsize \pageheight=\vsize \ruleht=.4pt
\abovedisplayskip=6pt plus 3pt minus 1pt
\belowdisplayskip=6pt plus 3pt minus 1pt
\abovedisplayshortskip=0pt plus 3pt
\belowdisplayshortskip=4pt plus 3pt

\newinsert\margin
\dimen\margin=\maxdimen




\newdimen\paperheight \paperheight = \vsize
\def\topmargin{\afterassignment\@finishtopmargin \dimen0}%
\def\@finishtopmargin{%
  \dimen2 = \voffset		
  \voffset = \dimen0 \advance\voffset by -1in
  \advance\dimen2 by -\voffset	
  \advance\vsize by \dimen2	
}%
\def\advancetopmargin{%
  \dimen0 = 0pt \afterassignment\@finishadvancetopmargin \advance\dimen0
}%
\def\@finishadvancetopmargin{%
  \advance\voffset by \dimen0
  \advance\vsize by -\dimen0
}%
\def\bottommargin{\afterassignment\@finishbottommargin \dimen0}%
\def\@finishbottommargin{%
  \@computebottommargin		
  \advance\dimen2 by -\dimen0	
  \advance\vsize by \dimen2	
}%
\def\advancebottommargin{%
  \dimen0 = 0pt\afterassignment\@finishadvancebottommargin \advance\dimen0
}%
\def\@finishadvancebottommargin{%
  \advance\vsize by -\dimen0
}%
\def\@computebottommargin{%
  \dimen2 = \paperheight	
  \advance\dimen2 by -\vsize	
  \advance\dimen2 by -\voffset	
  \advance\dimen2 by -1in	
}%
\newdimen\paperwidth \paperwidth = \hsize
\def\leftmargin{\afterassignment\@finishleftmargin \dimen0}%
\def\@finishleftmargin{%
  \dimen2 = \hoffset		
  \hoffset = \dimen0 \advance\hoffset by -1in
  \advance\dimen2 by -\hoffset	
  \advance\hsize by \dimen2	
}%
\def\advanceleftmargin{%
  \dimen0 = 0pt \afterassignment\@finishadvanceleftmargin \advance\dimen0
}%
\def\@finishadvanceleftmargin{%
  \advance\hoffset by \dimen0
  \advance\hsize by -\dimen0
}%
\def\rightmargin{\afterassignment\@finishrightmargin \dimen0}%
\def\@finishrightmargin{%
  \@computerightmargin		
  \advance\dimen2 by -\dimen0	
  \advance\hsize by \dimen2	
}%
\def\advancerightmargin{%
  \dimen0 = 0pt \afterassignment\@finishadvancerightmargin \advance\dimen0
}%
\def\@finishadvancerightmargin{%
  \advance\hsize by -\dimen0
}%
\def\@computerightmargin{%
  \dimen2 = \paperwidth		
  \advance\dimen2 by -\hsize	
  \advance\dimen2 by -\hoffset	
  \advance\dimen2 by -1in	
}%

\def\onepageout#1{\shipout\vbox{ 
    \offinterlineskip 
    \vbox to 3pc{ 
      \iftitle 
        \global\titlefalse 
        \setcornerrules 
      \else\ifodd\pageno \rightheadline\else\leftheadline\fi\fi
      \vfill} 
    \vbox to \pageheight{
      \ifvoid\margin\else 
        \rlap{\kern31pc\vbox to\z@{\kern4pt\box\margin \vss}}\fi
      #1 
      \ifvoid\footins\else 
        \vskip\skip\footins \kern-3pt
        \hrule height\ruleht width\pagewidth \kern-\ruleht \kern3pt
        \unvbox\footins\fi
      \boxmaxdepth=\maxdepth
      } 
    }
  \advancepageno}

\def\setcornerrules{\hbox to \pagewidth{\vrule width 1pc height\ruleht
    \hfil \vrule width 1pc}
  \hbox to \pagewidth{\llap{\sevenrm(page \folio)\kern1pc}%
    \vrule height1pc width\ruleht depth\z@
    \hfil \vrule width\ruleht depth\z@}}
\newbox\partialpage




\hyphenation{leut-wy-ler schil-cher}


\input epsf.sty
\def\wtightboxit{\relax}
\raggedbottom
\footline={\hfill}
\rightline{\timestamp}
\smallskip
\rightline{FTUAM  02-1}
\rightline{(hep-ph/xxx)}
\bigskip
\hrule height .3mm
\vskip.6cm
\centerline{{\twelverm Basic Parameters and 
Some Precision Tests of the Standard Model}\footnote{(*)}{
Talk given at the XXX International Meeting 
on Fundamental Physics, Jaca, January 28 -- February 1, 2002.}}
\medskip
\centerrule{.7cm}
\vskip1cm

\setbox9=\vbox{\hsize65mm {\noindent\fib 
 F. J. Yndur\'ain} 
\vskip .1cm
\noindent{\addressfont Departamento de F\'{\i}sica Te\'orica, C-XI,\hb
 Universidad Aut\'onoma de Madrid,\hb
 Canto Blanco,\hb
E-28049, Madrid, Spain.}\hb}
\smallskip
\centerline{\box9}
\vskip1truecm
\setbox0=\vbox{\abstracttype{Abstract} 
We present a review of the masses (except for neutrino masses)
 and interaction strengths in the standard model. 
Special emphasis is put on quantities that have been 
determined with significantly improved precision in the 
last few years. 
In particular, a number of determinations of $\alpha_s$ 
and the electromagnetic coupling on the $Z$, 
$\alpha_{\rm QED}(M_Z^2)$ are presented and their implications for  
 the Higgs mass  discussed; the best prediction 
that results for this last quantity being
$$M_H=102^{+54}_{-36}\;GeV/c^2.$$
Besides this, we also discuss a few extra precision tests of the 
standard model: the electron magnetic moment and dipole 
moment, and the muon magnetic moment. 
}
\centerline{\box0}
\brochureendcover{Typeset with \physmatex}
\pageno=1
\brochureb{\smallsc f. j.  yndur\'ain}{\smallsc 
basic parameters and some precision tests of the standard model}{1}

\booksection{1. Introduction}
The standard model (SM) of particle physics, based on the gauge interactions 
described by the group $SU(3)\times SU_L(2)\times U(1)$ 
supplemented by the Higgs-Weinberg mechanism for spontaneously breaking the symmetry,  
represents an incredibly precise and successful description of 
the microscopic Universe. Up to the highest energies explored by our accelerators, 
a few hundreds of \gev\ (per constituent, in the case of the Tevatron or HERA) 
and to an accuracy that reaches the 12 exact digits, for some electromagnetic 
processes, the standard model provides 
answers to essentially all the non-gravitational observed phenomena.\fnote{Of 
course, I am aware that this is not strictly true, as 
there are 
 questions which, while not contradicting the 
standard model, 
have also not been given a satisfactory answer 
in its context. I am referring to the 
existence of the Higgs particle: 
the observational evidence for its existence  
is not compelling; to the the question of 
the neutrino masses (it is unclear how to accommodate the nonzero masses 
in the model), and to the unexplained smallness of the 
strong CP violating phase, $\theta_{\rm QCD}$.} 
What is more, there is no established observation that 
contradicts the predictions of the SM: the scares that,  
from time to time, lead to hundreds of 
papers being written on possible {\sl new 
physics} signals, have invariably evaporated, and 
arguments (for example, of naturalness) about new physics 
lurking just around the bend are, at best, wishful thinking.

In this talk I will concentrate on the determination of 
some basic parameters;  to 
be precise, masses (except neutrino masses, 
already widely discussed at this Meeting) and couplings. 
I will also discuss  
 a few of the precision tests of 
the standard model, with special emphasis on 
those  to which I have had the occasion to contribute 
through some of the corresponding theoretical evaluations. 

\booksection{2. Masses and charges of electron, $\mu$ and $\tau$}
\vskip-0.5truecm
\booksubsection{2.1. Charge and mass of the electron. 
The electron magnetic moment}
When, in 1897, J.~J.~Thomson discovered the electron, he also tried to find its charge 
and mass. 
He could, however, not measure accurately enough the charge deposited in the anode. 
So, he only was able, by measuring the deviation of
 the electron in  Crooke's tube,  to determine the 
ratio $e/m$. This ratio was useful, for example, to 
identify the $\beta$ rays in radioactive decays of electrons 
(\fig~1). 
Nowadays, one measures independently the charge of the 
electron and, once this known, one derives its mass from the 
ratio $e/m$, measured by the bending of the 
trajectories of electrons in magnetic fields, just as Thomson and Chadwick did.

\bigskip
\setbox1=\vbox{\hsize7truecm{\epsfxsize=5.2truecm\epsfbox{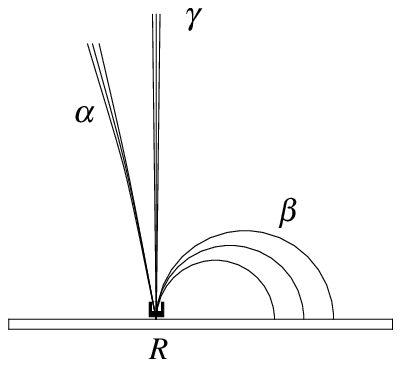}}}
\setbox2=\vbox{\hsize 5.5cm\figuresc{figure 1. }{The 
bending of the $\alpha$ and $\beta$ rays in a magnetic field 
allows identification of their nature.\hb
(According to a drawing of  Chadwick, 1921)
\hb
\phantom{X}
\hb
\phantom{X}}}
\line{{\wtightboxit{\box1}}\hfil\box2}
\bigskip

The more precise of the traditional methods for determining the charge of the electron are 
based on the Hall and Josephson effects.\ref{1} 
One finds, for $\alpha=e^2/4\pi\hbar c$,

$$\eqalign{
\alpha=&\dfrac{1}{137.035\,977\,0\;[77]}\qquad\hbox{[Josephson effect]}\cr
\alpha=&\dfrac{1}{137.036\,003\,7\;[33]}\qquad\hbox{[Quantum Hall effect]}.\cr}
\equn{(2.1)}
$$
Note that these numbers differ by more than three standard deviations. 
The more precise and independent determination of $\alpha$ from 
the electron magnetic moment, to be discussed presently, 
favours the value obtained from the quantum Hall effect.

The more precise value for $\alpha$, as 
anticipated, is an indirect one.
Consider the anomalous magnetic moment of the electron, $a({\rm e})$. 
To fourth order in electromagnetic interactions, we can write\ref{1}
$$\eqalign{a({\rm e})=&\tfrac{1}{2}\left(\dfrac{\alpha}{\pi}\right)+
c_2\left(\dfrac{\alpha}{\pi}\right)^2
+c_3\left(\dfrac{\alpha}{\pi}\right)^2+c_4\left(\dfrac{\alpha}{\pi}\right)^2;\cr
c_2=&-0.328\,478\,965,\quad c_3\simeq1.181\,241\,456,\quad c_4\simeq-1.610\pm0.038.\cr}
$$
(Actually, $c_2$ and $c_3$ are known exactly, the analytical 
evaluation of last completed very recently;
 only for $c_4$ 
there exists only  numerical evaluations). 
Including also the very small [$(4.393\pm0.027)\times10^{-12}$] 
$\mu$ and $\tau$ loop and weak and strong corrections one has, using (2.1), 
and comparing with experiment, 
$$\eqalign{a({\rm e})=&\big(1\,159\,652\,140\pm 27\big)
\times 10^{-12}\quad\hbox{[Theory]},\cr
a({\rm e})=&\big(1\,159\,652\,188\pm 3\big)\times 10^{-12}\quad\hbox{[Experimental].}
\cr}
\equn{(2.2)}$$
The agreement is astonishing: eleven significant digits! 
Indeed, one should remember that the magnetic moment is 
given by the formula
$$\mu_{\rm e}=[1+a({\rm e})]\mu_{\rm B}$$
where $\mu_{\rm B}$ is the Bohr magneton, given by the 
Dirac equation: theory 
predicts the whole of $\mu_{\rm e}$, not merely the anomaly.
 
What is perhaps even more impressive is that most of the 
theoretical error comes from the {\sl experimental} error in $\alpha$, \equn{(2.1)}. 
This suggests that we reverse the argument and deduce $\alpha$ from $a({\rm e})$. 
If we do this, we get 
$$\alpha=
\dfrac{1}{137.035\,999\,6\;[5]}\qquad \hbox{[Deduced from $a({\rm e})$]},
\equn{(2.3)}$$
almost six times more precise than the values obtained with the 
traditional 
methods based on the Hall and Josephson effects.

Because of radiative corrections, the intensity 
of the interactions depend on the energy at which they are measured. 
The value we have recorded for $\alpha$ is that measured at 
zero energy. 
In many applications one needs the value of $\alpha$ at the rest energy of the 
$Z$ particle, $\alpha(M^2_Z)$. 
The more recent figure for this quantity is\ref{2}
$$\alpha(M_Z^2)=\dfrac{1}{128.965\;[17]}.
\equn{(2.4)}$$

Once $\alpha$ known one can 
obtain the mass of the electron from the ratio $e/m_e$. 
According to the Particle Data Group (PDG\ref{3}) one has
$$m_e=0.510\,999\,05[15]\quad \mev/c^2.
\equn{(2.5)}$$

\booksubsection{2.2. Charge and masses of $\mu$, $\tau$}
The identity of the charges of the electron, $\mu$ ad $\tau$ follow 
from charge conservation,  implied by the masslessness of the 
photon. 
This is because of the existence of the decays
$$\mu\to\bar{\nu}\nu e,\quad \tau\to \bar{\nu}\nu l,\;l=e,\;\mu$$
and the zero electric charge of the neutrinos,\fnote{That the 
charge of the neutrinos must be zero or at least extremely small 
follows from the fact that solar $\nu$'s arrive on Earth. 
These neutrinos are produced at the center of the 
Sun and, if they had the minutest charge, they could not have 
escaped  traversing the whole of the Sun.}
very well established.

The masslesness of the photon, in turn, 
is linked to the range of electromagnetic interactions. 
Since these have been measured at some $200\,000$~Km from the source 
(the magnetic field of Jupiter), we have a bound on the photon mass of
$$m_\gamma\leq 10^{-32}\;\mev/c^2.
$$  
In view of this, we are justified in assuming the charges of $\mu$, $\tau$ to be  
identical to that of the electron.

As for the masses of these particles, that of the $\tau$ follows from 
measurements of the location of  
the threshold for $\tau^+\tau^-$ production at the Beijing collider. 
The PDG\ref{3} give the figure
$$m_\tau=1777.03^{+0.30}_{-0.26}\;\mev/c^2.
\equn{(2.6)}$$

The $\mu$ particle lives long enough (a millionth of a second) 
that it leaves clear tracks, and hence its mass 
can be obtained (since its charge is known) as for the electron, measuring the curvature of 
these tracks. This way one gets,\ref{3}
$$m_\mu=105.658389\pm0.000034\;\mev/c^2.
\equn{(2.7)}$$
However, and as for the electron charge, 
an indirect method exists that gives a more precise 
value. 
Consider the hyperfine splitting, $\Delta\nu$, in 
the hydrogen-like bound state $(\mu^+e^-)$. 
It can be measured experimentally, and evaluated theoretically. 
The (very complicated) theoretical calculation is discussed 
in the contribution of J.~Sapiristein and D.~Yennie in ref.~4.
One has,
$$\eqalign{\Delta \nu=\tfrac{8}{3}\dfrac{m_r\alpha^3}{m_em_\mu}\left[1+a({\mu})\right]
&\Bigg\{1+a({\rm e})+\left(\log2-1\right)\alpha^2
-\left(\log\alpha-2\log2+\tfrac{281}{480}\right)\dfrac{8\log\alpha}{3\pi}\alpha^3
+D\alpha^3\cr
-&\dfrac{3\alpha}{\pi}\dfrac{m_e m_\mu \log(m_\mu/m_e)}{m^2_\mu-m^2_e}
+\dfrac{m_e\alpha^2}{m_\mu}\left[2\log\dfrac{m_r}{2\gamma}-6\log2+\tfrac{11}{6}\right]\cr
+&\left(\dfrac{\alpha^2}{\pi}\right)\dfrac{m_e}{m_\mu}
\left[-2\log^2\dfrac{m_\mu}{m_e}+\tfrac{13}{12}\log\dfrac{m_\mu}{m_e} 
+\tfrac{21}{2}\zeta(3)+\dfrac{\pi^2}{6}+\tfrac{35}{9}+ 
(1.9\pm0.3)\right]
\Bigg\}.
\cr}$$
Here $m_r$ is the reduced mass, $m_r^{-1}=m_e^{-1}+m_{\mu}^{-1}$, $\gamma=m_e\alpha$ 
and $D=(15.38\pm0.29)/\pi+D_1$, with $D_1$  an as yet uncalculated constant 
related to diagrams involving 
two virtual photons; see the quoted article of Sapiristein and Yennie.
If we took the value of $m_\mu$ from direct measurements, (2.7), we would find\ref{4}
$$\eqalign{
\Delta \nu=&4\,463\,303.11\pm2.6\quad\hbox{[Theory]}\cr
\Delta \nu=&4\,463\,302.88\pm0.16\quad\hbox{[Experiment]}.\cr}
\equn{(2.8)}$$
The error in the theoretical evaluation is due almost exclusively to 
that of $m_\mu$ from direct measurements. Hence, as with the anomalous magnetic moment of the 
electron and its charge, we may reverse the argument and obtain, 
from the experimental hyperfine splitting,  $\Delta\nu$, and 
the theoretic calculations, a figure for the muon mass 
seven times more precise than the value obtained with traditional methods:
$$m_\mu=105.658357\pm0.000005\;\mev/c^2.
\equn{(2.9)}$$

\booksection{3. The strong interaction coupling, $\alpha_s$, and quark masses}
\vskip-0.5truecm
\booksubsection{3.1.  The strong interaction coupling}
There are, in the recent years, a number of theoretical 
calculations that have been carried to NNLO (next to next to leading order) in 
$\alpha_s$. This has permitted an evaluation of 
$\alpha_s$ to an error of 1\%; the consistency of the 
calculations furnishes a set of tests of QCD at the 
level of a few percent. 
The processes in question are

\item{1. }{Hadronic decays of the $\tau$.}
\item{2. }{Hadronic decays of the  $Z$.}
\item{3. }{Hadronic annihilations of $e^+e^-$.}
\item{4. }{Deep inelastic collisions of electrons, muons and neutrinos with nucleons 
(DIS).}
In addition we have sum rules, in particular the Bjorken and 
Gross--Llewelyn~Smith sum rules.

\topinsert
{\phantom{X}
\vskip0.5truecm
\medskip
\setbox0=\vbox{\hsize 9truecm \epsfxsize=7.5truecm\epsfbox{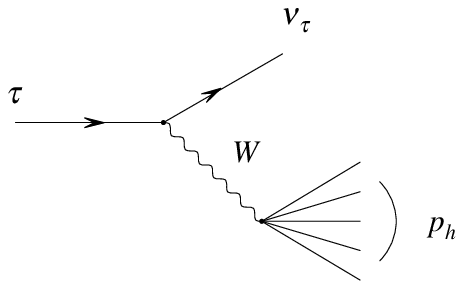}}
\setbox1=\vbox{\hsize 9.8truecm \epsfxsize=8.1truecm\epsfbox{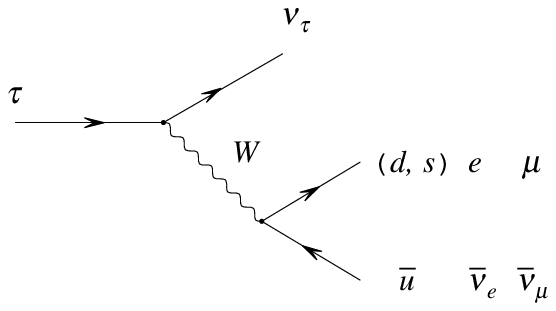}}
\setbox2=\vbox{\hsize 5.5truecm
 \captiontype\figuresc{figure 2. }{Hadronic decay of the  $\tau$, 
and scheme of the decay   
into leptons and quarks.}\hb
\vskip 1truecm
\phantom{X}}
\setbox9=\vbox{\hsize 10 truecm
\centerline{\wtightboxit{\box0}}
\centerline{\wtightboxit{\box1}}}
\line{\box9\hfil\box2}
\bigskip
}\endinsert
\medskip 
The ratio of the decays of  $\tau$ into hadrons and leptons (Fig.~2), and the like ratios  
for $Z$ decays and $e^+e^-$ annihilations
$$R=\dfrac{\sigma(e^+e^-\to \hbox{hadrons})}{\sigma(e^+e^-\to\mu^+\mu^-},$$ 
yielded some of the first precise measurements of $\alpha_s$, 
because they were the first to be evaluated to NNLO. 
For example, one has, with  $s^{1/2}$ the c.m. energy, 
and neglecting quark masses,\ref{5}
$$R(s)=3\sum_{f=1}^{n_f}Q^2_f\left\{1+\dfrac{\alpha_s(s)}{\pi}+
r_2\left(\dfrac{\alpha_s(s)}{\pi}\right)^2
+r_3\left(\dfrac{\alpha_s(s)}{\pi}\right)^3\right\}+O(\alpha_s^4),
\equn{(3.1a)}$$
where 
$$\eqalign{
r_2=&\tfrac{365}{24}-11\zeta(3)+\left[\tfrac{2}{3}\zeta(3)-\tfrac{11}{12}\right]n_f
\simeq2.0-0.12n_f.\cr
r_3=&-6.637-1.200n_f-0.005n_f^2-1.240\left(\sum_1^{n_f}Q_f\right)^2
\left(3\sum_1^{n_f}Q_f^2\right)^{-1}.
\cr}
\equn{(3.1b)}$$
$Q_f$ are the quark charges, and $n_f$ the number of active flavours. 
The three loop expression for the running constant
 has to be used here:\fnote{Actually, $\alpha_s$ is known to four loops; 
cf. ref.~6.}
$$\eqalign{\alpha_s(Q^2)=&\dfrac{4\pi}{\beta_0L}\left\{1-\dfrac{\beta_1\log L}{\beta_0^2L}+
\dfrac{\beta_1^2\log^2L-\beta_1^2\log L+
\beta_2\beta_0-\beta_1^2}{\beta_0^4L^2}\right\},
\cr
L=&\log Q^2/\Lambdav^2.\cr}$$
The values of the $\beta_i$ are given after \equn{(3.7)}.

DIS (\fig~3) furnished the first NLO evaluations of $\alpha_s$, and now 
provide the more precise NNLO results, because their energy range is so large.
Moreover, they also give an important check of QCD in that the 
theoretical evaluations successfully predict the experimental scaling violations.

\topinsert
{
\setbox1=\vbox{\hsize 8.5truecm \epsfxsize=6.7truecm\epsfbox{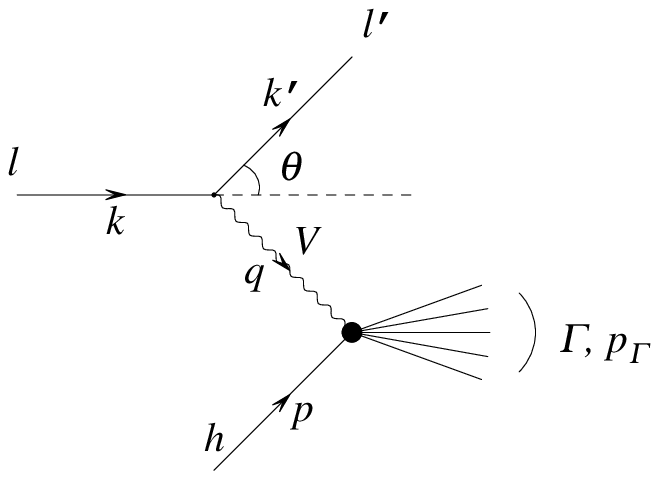}}
\setbox2=\vbox{\hsize 5.truecm
 \captiontype\figuresc{figure 3. }{Deep inelastic scattering of leptons, $l$, $l'$, 
with a hadron  
 $h$. The virtual particle  $V$ may be a photon, $W$ or $Z$.}
\hb
\vskip1truecm
\phantom{X}}
\line{\wtightboxit{\box1}\hfil\box2}
}\endinsert

The results are summarized in Table~1. 
There,  DIS means deep inelastic scattering, 
Bj stands for the Bjorken, and GLS for the Gross--Llewellyn Smith 
sum rules.\ref{6} 
The values given for $Z\to \hbox{hadrons}$\ref{8} assume the Higgs mass 
constrained by $100\leq M_H\leq 200\;\gev$, with the central value for $M_H=115\;\gev$.
We have presented two results here for $e/\mu p$ DIS,\ref{7}
 the one for the smaller $Q^2$ range (I, probably 
 the more reliable) and the extended range one, labeled~II. 
DIS provides the more precise determination of 
$\alpha_s$, which is not surprising: the 
extended $Q^2$ range allows a sizable variation to the 
logarithmic dependence of $\alpha_s$ on $Q^2$.

\setbox0=\vbox{\petit
\medskip
\setbox1=\vbox{\offinterlineskip\hrule
\halign{
&\vrule#&\strut\hfil#\hfil&\quad\vrule\quad#&
\strut\quad#\quad&\quad\vrule#&\strut\quad#\cr
 height2mm&\omit&&\omit&&\omit&\cr 
& \kern.5em Process&&${\textstyle\hbox{Average}\;
 Q^2}\atop{\textstyle \hbox{or}\; Q^2\;\hbox{range}\;[\gev]^2}$&
& $\alpha_s(M_Z^2)$\kern.3em& \cr
 height1mm&\omit&&\omit&&\omit&\cr
\noalign{\hrule} 
height1mm&\omit&&\omit&&\omit&\cr
&\phantom{\Big|}  DIS; $\nu$, Bj&&1.58&&$0.121^{+0.005}_{-0.009}$\phantom{l}& \cr
\noalign{\hrule}
&\phantom{\Big|}  DIS; $\nu$, GLS&&3&&$0.112\pm0.010$\phantom{l}& \cr
\noalign{\hrule}
&\phantom{\Big|}  $\tau$ decays&&$(1.777)^2$&&$0.1181\pm0.0031$&\cr
\noalign{\hrule}
&\phantom{\Big|}  $e^+e^-\to{\rm hadrons}$&&$100 - 1600$&&$0.128\pm0.025$&\cr
\noalign{\hrule}
&\phantom{\Big|}  $Z\to{\rm hadrons}; \Gammav_Z$&&$(91.2)^2$&&$0.1230\pm0.0038$&\cr 
\noalign{\hrule}
&\phantom{\Big|}  $Z\to{\rm hadrons}; \hbox{GrandLEP}$&&$(91.2)^2$&&$0.1185\pm0.0030$&\cr 
\noalign{\hrule}
&\phantom{\Big|}  $Z\to{\rm jets}$&&$(91.2)^2$&&0.117&\cr
\noalign{\hrule}
&\phantom{\Big|}  DIS($\nu\,N;\;xF_3$)&&$8 - 120$&&$0.1153\pm0.0041$\phantom{l}& \cr
\noalign{\hrule}
&\phantom{\Big|}  DIS ($ep$, I)&&${3.5} - {230}$&&$0.1166\pm0.0013$\phantom{l}&\cr
\noalign{\hrule}
&\phantom{\Big|}  DIS ($ep$, II)&&${3.5} - {5000}$&&$0.1163\pm0.0014$\phantom{l}&\cr
 height1mm&\omit&&\omit&&\omit&\cr
\noalign{\hrule}}
\vskip.05cm}
\centerline{\box1}
\smallskip
\centerline{\petit Table 1}
\centerrule{6cm}
\medskip}
\box0

As is seen, the more precise evaluations use DIS for $ep$ scattering. 
These in turn employ the calculations for the 
so-called Wilson coefficients and anomalous dimensions. 
Of these, only the ones relating to a few moments (six- seven) are known. 
This number increased in the last year; before, only four 
were known. 
Using only these four, the value obtained would be
$$\alpha_s(M_Z^2)=0.1172\pm0.0024 \quad ep,\;\hbox{(DIS, only four moments)}.$$
The consistency with the values quoted in the Table is a proof of the stability of the
DIS  calculations.

  All determination are compatible with 
one another, within errors. The world average is 

$$\alpha_s(M_Z^2)=0.1173\pm 0.0011.
\equn{(3.2)}$$

\booksubsection{3.2.The mass of the $t$ quark}
The quark $t$ was predicted theoretically long before it was 
found experimentally. In 1977, Veltman 
showed, from an analysis of radiative corrections,
 that the measurements of the parameters of electroweak interactions 
could be made compatible only if 
 $m_t$ was less than some $300\;\gev/c^2$. 
Then the measurements of those parameters were refined, especially after the 
beginning of operations of LEP, and the error in the theoretical 
prediction improved to 10\%, which was the value just prior to 
experimental discovery. 
It, of course, has gone on improving, as we will show presently.

The reason why one gets so sensitive results 
is that the $\rho$-parameter 
(essentially, the ratio of the $Z$, $W$ masses) 
depends quadratically on $m_t$.\ref{9} To one loop, 
$$M_W/M_Z=\left\{1+\dfrac{3G_Fm^2_t}{8\sqrt{2}\pi^2}\right\}\cos \theta_W.$$
$\theta_W$ is called the weak mixing angle, and measures the mixing of 
electromagnetic and neutral weak currents. 
$G_F$ is the Fermi constant, given below.

Presently, we have two evaluations of the mass of the quark $t$. 
The mass deduced from observables other than 
the $t$ quark itself is, for the pole mass, 
$$m_t=168.2^{+9.6}_{-7.4}\gev/c^2\quad\hbox{[Radiative corrections]}
\equn{(3.3)}$$
while the direct experimental value, obtained after seven years of measurements, is
$$m_t=174.3\pm5.1\gev/c^2\quad\hbox{[Tevatron]}.
\equn{(3.4)}$$
The agreement is more impressive if we realize that 
part of the theoretical error is due to the 
uncertainty in the mass of the  Higgs particle.

We have taken both values from the PDG.\ref{3} Note that 
the mass obtained with the direct measurement can only be identified 
with the pole mass if one neglects the decay of the $t$, 
so (3.4) should have an extra error, likely small, of 
 order $m_t\alpha_s\alpha_W\sim 0.6\;\gev$.\fnote{I am grateful
 to P.~Langacker for discussions and 
information on this.}

\booksubsection{3.3. The masses of the quarks $b$, $c$}
The more precise evaluations of the masses of these quarks are those obtained from quarkonia 
states. 
Here the precision is of $O(\alpha_s^4)$ and the leading nonperturbative 
correction, involving the gluon condensate $\langle\alpha_sG^2\rangle_{\rm vac.}$ 
is also incorporated. 
For the $b$ quark mass, the corrections due to the finite
 mass of the $c$ quark, and the leading correction of 
order  $O(\alpha_s^5\log\alpha_s)$ can also be included. 
The subleading nonperturbative corrections can be estimated, and are (relatively) small. 
Thus, we have a very precise evaluation of the mass of this quark. 
For the $c$ quark, however, the nominally subleading 
corrections are so large that it is not even clear to what extent 
the inclusion of the perturbative corrections of  $O(\alpha_s^4)$ 
improve the results.

The  mass of the heavy $b$, $c$ quarks may then be obtained by solving the
 implicit equation that gives, in terms of it,  the mass 
of the ground state (for which one sets $n=1,\,l=0$ below) 
of the corresponding quarkonium state,
$$E_{nl}=2m-m\dfrac{C_F^2\widetilde{\alpha}_s^2}{4n^2}+\sum_{V}\delta^{(1)}_{V}E_{nl}
+\delta^{(2)}_{V_1^{(L)}}E_{nl}+\delta_{\rm NP}E_{nl}.\equn{(3.5)}$$
We define the analogue of the Bohr radius,
$$a(\mu^2)=\dfrac{2}{mC_F\widetilde{\alpha}_s(\mu^2)},$$
and one has\fnote{The one loop static correction 
can be found in ref.~10; the velocity dependent one in ref.~11. 
Two loop corrections are from ref.~12, 
and the rest of the $O(\alpha_s^4)$ contributions are from 
ref.~13.}
$$\eqalign{
\tilde{\alpha}_s(\mu^2)=&\alpha_s(\mu^2)
\left\{1+\left(a_1+\dfrac{\gammae\beta_0}{2}\right)\dfrac{\alpha_s(\mu^2)}{\pi}\right.\cr
&\left.+\left[\gammae\left(a_1\beta_0+\dfrac{\beta_1}{8}\right)+
\left(\dfrac{\pi^2}{12}+\gammae^2\right)\dfrac{\beta_0^2}{4}+
b_1\right]\dfrac{\alpha_s^2}{\pi^2}\right\};\cr}\equn{(3.6a)}$$
$$\eqalign{a_1=&\dfrac{31C_A-20T_Fn_f}{36}\simeq 1.47;\cr
b_1=&\tfrac{1}{16}
\Big\{\left[\tfrac{4343}{162}+4\pi^2-\tfrac{1}{4}\pi^4+\tfrac{22}{3}\zeta(3)\right]C_A^2\cr
&-\left[\tfrac{1798}{81}+\tfrac{56}{3}\zeta(3)\right]C_AT_Fn_f-
\left[\tfrac{55}{3}-16\zeta(3)\right]C_FT_Fn_f+\tfrac{400}{81}T_F^2n_f^2\Big\}\cr
&\simeq 13.2. \cr}\equn{(3.6b)} $$
Moreover, the sum over $V$ in (3.5) runs over the following pieces:
$$\delta^{(1)}_{V_{\rm tree}}E_{nl}=-\dfrac{2}{n^3\,m^3\,a^4}
\left[\dfrac{1}{2l+1}-\dfrac{3}{8n}\right]+
\dfrac{C_F\alpha_s}{m^2}\,\dfrac{2l+1-4n}{n^4(2l+1)a^3};\equn{(3.7a)}$$
$$\delta^{(1)}_{V^{(L)}_1}E_{nl}=
-\dfrac{\beta_0C_F\alpha^2_s(\mu^2)}{2\pi n^2a}
\left[\log\dfrac{na\mu}{2}+\psi(n+l+1)\right];\equn{(3.7b)}$$
$$\delta^{(1)}_{V_2^{(L)}}E_{nl}=-\dfrac{C_Fc_2^{(L)}\alpha_s^3}{\pi^2n^2a}\;
\left[\log\dfrac{na\mu}{2}+\psi(n+l+1)\right];\equn{(3.7c)}$$
$$\eqalign{\delta^{(1)}_{V^{(LL)}}E_{nl}=-\dfrac{C_F\beta_0^2\alpha_s^3}{4\pi^2n^2a}\,
\Big\{\log^2\dfrac{na\mu}{2}+2\psi(n+l+1)\log\dfrac{na\mu}{2}\cr
+\psi(n+l+1)^2+\psi'(n+l+1)\cr
+\theta(n-l-2)\dfrac{2\Gammav(n-l)}{\Gammav(n+l+1)}
\sum^{n-l-2}_{j=0}\dfrac{\Gammav(2l+2+j)}{j!(n-l-j-1)^2}\Big\};\cr}
\equn{(3.7d)}$$
$$\delta^{(1)}_{V_{\rm s.rel}}E_{nl}=\dfrac{C_Fa_2\alpha_s^2}{m}\;
\dfrac{1}{n^3(2l+1)a^2}\equn{(3.7e)}$$
and 
$$\delta^{(2)}_{V_1^{(L)}}E_{nl}\equiv 
-m\dfrac{C_F^2\beta_0^2\alpha_s^4}{4n^2\pi^2}
\left\{N_0^{(n,l)}+N_1^{(n,l)}\log \dfrac{na\mu}{2}+
\tfrac{1}{4}\log^2\dfrac{na\mu}{2}\right\}. \equn{(3.7f)}$$
Here, 
$$\eqalign{N_1^{(1,0)} = -{\gamma_E \over 2} \simeq -0.288608 \cr
N_1^{(2,0)} = {1 -2 \gamma_E \over 4} \simeq -0.0386078 \cr
N_1^{(2,1)} = {5 -6 \gamma_E \over 12} \simeq 0.128059 \cr
N_0^{(1,0)}= {3 +3 \gamma_E^2 -\pi^2 +6 \zeta(3) \over 12} \simeq 0.111856 \cr
N_0^{(2,0)}=-\tfrac{5}{16}-{\gamma_E \over 4}+{\gamma_E^2 \over 4} - 
{\pi^2 \over 12} + \zeta(3) \simeq 0.00608043 \cr
N_0^{(2,1)}=-\tfrac{865}{432}-{5 \gamma_E \over 12}+{\gamma_E^2 \over 4} - 
{11\pi^2 \over 36} + \zeta(3) \simeq 0.0314472, \hbox{etc.} \cr}
$$
For the vector states ($\Upsilonv,\,\Upsilonv',\,\Upsilonv'';\;J/\psi,\,\psi',\dots$) 
one has to add the hyperfine shift, at tree level (setting $s$, the spin, equal to unity):
$$\delta^{(1)}_{V_{\rm spin}}E_{nl}=\delta_{s1}\delta_{l0}
\dfrac{8C_F\alpha_s}{3n^3m^2a^3}.
\equn{(3.7g)}$$
In these formulas,
$$\eqalign{\beta_0=&11-\tfrac{2}{3}n_f;\beta_1=102-\tfrac{38}{3}n_f\cr
\beta_2=&\tfrac{2847}{2}-\tfrac{5033}{18}n_f+\tfrac{325}{54}n_f^2\cr
c_2^{(L)}=&a_1\beta_0+\tfrac{1}{8}\beta_1+\tfrac{1}{2}\gammae\beta_0^2,\cr
C_F=&4/3,\quad C_A=3,\quad T_F=1/2,\cr}$$
and $\mu$ is a reference momentum, usually taken 
to be $\mu=2/na$.

In addition to this one has to consider the nonperturbative (NP) 
contributions to the energy levels. 
The dominant ones are associated with the gluon condensate and are\ref{14}
$$\eqalign{\delta_{\rm NP}E_{nl}=
m\epsilon_{nl}n^2\pi\langle\alpha_sG^2\rangle\left(\dfrac{na}{2}\right)^4=
m\dfrac{\epsilon_{nl}n^6\pi\langle\alpha_sG^2\rangle}{(mC_F\tilde{\alpha}_s)^4};\cr
\epsilon_{10}=\tfrac{1\,872}{1\,275},\;\epsilon_{20}=\tfrac{2\,102}{1\,326},\;
\epsilon_{21}=\tfrac{9\,929}{9\,945}.\cr}\equn{(3.8)}$$

We will not give explicit formulas for the subleading nonperturbative effects, the 
  $O(\alpha_s^5\log\alpha_s)$ corrections\ref{15} or the finite $c$ mass contribution\ref{16}
 (for the 
$b$ quark). The result is, taking them into account,\ref{16a}  
$$m_b=5\,022\pm 58\;\mev/c^2;\quad
\bar{m}_b(\bar{m}_b^2)=4\,285\pm 36\;\mev/c^2.
\equn{(3.9)}$$
The relation between the \msbar\ mass and the pole mass is evaluated taking into account 
one, two and three loop corrections,\ref{17} so, while $m_b$ is exact to 
  $O(\alpha_s^5\log\alpha_s)$, $\bar{m}_b(\bar{m}_b^2)$ is ``only" 
correct to $O(\alpha_s^3)$.

For the $c$ quark we have
$$\eqalign{
m_c=&1\,866^{+215}_{-133}\;\mev/c^2;\quad O(\alpha_s^4)\cr
m_c=&1\,570\pm100\;\mev/c^2;\quad O(\alpha_s^3)+O(v^2).\cr}
\equn{(3.10a)}$$
Here the first result is that of ref.~13, the second from ref.~11. 
As stated, it is unclear  which is more believable. 
The reason is that adding two loop corrections shifts the mass beyond the 
errors that are obtained for the one loop result. 
The corresponding values of the \msbar\ mass are,
$$\eqalign{
\bar{m}_c=&1\,542^{+163}_{-104}\;\mev/c^2\cr
\bar{m}_c=&1\,306\pm90\;\mev/c^2\cr}
\equn{(3.10b)}$$

Other methods for estimating the masses use the decay $Z\to\bar{b}b+G$, 
lattice calculations or sum rules.\fnote{For a summary see ref.~18.} 
While giving results compatible with the former, they are less precise.

We finish this subsection with a few words to clarify a matter which 
is at times misunderstood. 
Evaluating the $E_{nl}$ as we have done here is {\sl not} 
a model calculation; no assumptions about ``confining potentials" or the like are 
made. The method of solving the Schr\"odinger equation 
with a static potential obtained from a field-theoretic calculation 
to a fixed order $N$ (two loops, in our case) and 
perturbing with relativistic and other corrections is indeed 
fully equivalent to a field theoretic evaluation, in which one 
rearranges the 
perturbative series adding infinite ladders to a kernel, obtained 
at order $N$, as in \fig~4.

\topinsert{
\setbox0=\vbox{\hsize 10truecm\epsfxsize8.5truecm\epsfbox{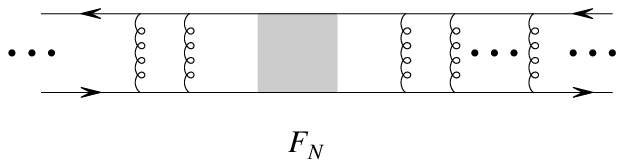}\hfil}
\setbox1=\vbox{\hsize 11.5truecm\captiontype\figuresc{figure 4. }
{``Ladders" at both sides of the  
{\sl kernel} $F_N$, to evaluate the $S$ matrix, and hence the bound state spectrum, 
for $\bar{q}q$.}
\vskip.8cm}
\centerline{\wtightboxit{\box0}}
\medskip
\centerline{\box1}}
\endinsert
\bigskip

\booksubsection{3.4. The masses of the light quarks: $u,\,d,\,s$}
Light quark masses are smaller than  the QCD parameter $\lambdav$, so 
bound state evaluations are useless to get them. 
One may use lattice calculations, or chiral 
perturbation arguments (for the 
ratios) and chiral sum rules,\ref{19,20} for the values of specific combinations 
of masses. 
Only bounds may be obtained from first principles; 
to get specific values models are needed.
  
In these methods one considers the pseudoscalar condensate,
$$\Psiv^{12}_5(t)=\ii \int\dd^4x\langle{\rm vac}|T\partial^{\mu}A^{ud}_{\mu}(x)
\partial^{\nu}A^{ud}_{\nu}(0)^{\dag}|{\rm vac}\rangle,\equn{(3.11a)}$$
$t=-q^2$ and $|{\rm vac}\rangle$ is the physical vacuum; the axial current 
is
$$A^{ud}_{\mu}(x)=\bar{q}_u(x)\gamma_{\mu}\gamma_5q_d(x),\equn{(3.11b)}$$ 
(for, e.g., the $ud$ combination). 
Then one writes the dispersion relation
$$\eqalign{F^{12}_5(t)=&
\int^{\infty}_0\dd s\,\dfrac{1}{(s+t)^3}\,\dfrac{2\imag \Psiv^{12}_5(s)}{\pi};\cr
F^{12}_5(t)=&\partial^2 \Psiv^{12}_5(t)/\partial t^2.\cr}$$
One can then calculate the l.h.s. with QCD, at large $t$;  to leading order,\fnote{For 
higher order corrections, see ref.~21.}
$$F^{12}_5(t)\eqsub_{\rm LO}\dfrac{3}{8\pi^2}\dfrac{[m_u(t)+m_d(t)]^2}{t},
\;t\gg\Lambdav^2,$$
 and the r.h.s. is evaluated  
with the pion contribution, for small $s$ and, for large $s$, 
again with QCD. 
Then one either 
 writes an inequality, noting that in the remaining 
intermediate energy region $\imag \Psiv^{12}_5(s)\geq0$, 
or one approximates $\imag \Psiv^{12}_5(s)$ for intermediate $s$  by the $\pi'$ pole.
In this way, we get two types of results: the rigorous, 
model independent bounds 
$$\bar{m}_d(1\,\gev)+\bar{m}_u(1\,\gev)\geq 9\,\mev/c^2,
\quad \bar{m}_d-\bar{m}_u\geq3\,\mev/c^2,
\quad \bar{m}_s(1\,\gev)\geq150\,\mev/c^2
\equn{(3.12)}$$
or the (somewhat model dependent)  absolute values
$$\eqalign{\bar{m}_d(1\,\gev)&+\bar{m}_u(1\,\gev)=13\pm4\,\mev/c^2,\cr
\bar{m}_s(1\,\gev)&=200\pm50\,\mev/c^2,\; \bar{m}_d(1\,\gev)=8.9\pm4.3\,\mev/c^2,
\; \bar{m}_u(1\,\gev)=4.2\pm2\,\mev/c^2.\cr}
\equn{(3.13)}$$
 
From decays into strange particles of the $\tau$ 
one also obtains a precise determination of the mass of the $s$ quark:\ref{22}
$$m_s(1\,\gev)=235^{+35}_{-42}\;\mev/c^2,
\equn{(3.14)}$$
in agreement with the former results, but with the advantage that it 
does not require the use of models.

Lattice results are compatible with these, but less reliable in that the 
quenched approximation (necessary to obtain 
meaningful calculations) is not well justified for the corresponding evaluations. 
Indeed, at times lattice results are incompatible with the values reported above,  
even with the bounds (3.12). 
The values of the light quark masses 
obtained with these other methods are discussed in 
some detail in the text of the author, ref.~18.

\booksection{4. Masses and interaction intensity for weak interactions}
\vskip-0.5truecm
\booksubsection{4.1. $\alpha_W$ and $G_F$}
Traditionally, instead of the 
intensity of the weak interaction, 
$\alpha_W\simeq 0.034$, one gives the  precision value for the Fermi coupling, $G_F$, 
linked to the former via the $W$ mass by the relation (to first order)
$$G_F=\dfrac{\pi}{\sqrt{2}M^2_W}\,\alpha_W.$$
According to the PDG\ref{3},
$$
G_F=1.16639[1]\times 10^{-5}\;\gev^{-2}.
\equn{(4.1)}$$

\booksubsection{4.2. $M_W$, $M_Z$ and the mass of the Higgs particle, $M_H$}
As for the masses of the $W$ and $Z$ particles the precision measurements of 
LEP and Tevatron give (PDG, ref.~3)
 $$\eqalign{M_W=&80.419\pm0.056\;\gev/c^2,\cr
M_Z=&91.1882\pm0.0022\;\gev/c^2.\cr}
\equn{(4.2)}$$
From these masses, the value of $\alpha(M_Z)$, and $G_F$, it is 
possible to deduce the mass of the Higgs 
particle by consistency. 
Using the LEPEWWG method one finds,  
$$M_H=102^{+54}_{-36}\;\gev/c^2\quad\hbox{[Radiative corrections]} 
\equn{(4.3)}$$
(Gr\"unwald, private communication). 
A similar (but much less precise) value follows from consistency of determinations of $\alpha_s$ 
(M.~J.~Herrero and the author, unpublished).
One may take the value of this quantity obtained from processes 
involving energies much smaller than $91.2\;\gev$, and extrapolate to 
predict $\alpha_s(M_Z^2)$. 
Comparing this with measurements of  $\alpha_s(M_Z^2)$ on the 
$Z$ itself (see the Table in  \sect~3.1) one gets consistency at $1.5\;\sigma$ level only if 
$M_H\leq133\;\gev$, and at $2\;\sigma$ if $M_H\leq493\;\gev$.

In the months
 prior to the decomissioning of LEP, evidence 
 was found for this particle with a mass of  
$115\;\gev/c^2$. 
The same experiments produced the bound $M_H\geq 114\;\gev$.
While the evidence is inconclusive, the coincidence with the value found from 
radiative corrections, (4.3), is encouraging. 

\booksection{5. A note on the magnetic moment of the muon, and the dipole moment of the electron}
In this last section we present two precision tests of the 
standard model, beyond those afforded by 
the consistencies already noted. 
The first is an extension of that already discussed, the magnetic moment of the 
muon. 
Besides its recent popularity, this is interesting because it involves all three interactions. 
The second is the dipole moment of the electron, that furnishes the best lower 
bound on new physics.

\booksubsection{5.1. The muon anomaly}
We have in \subsect~2.2 mentioned the equality of the $e$, $\mu$ charges. 
This is, in particular, a consequence of the structure of the standard model, 
especially of so-called electron-muon 
universality that specifies that all properties, 
except the mass, of $e$ and $\mu$ are identical. 
As a consequence of this it follows that we can evaluate the 
magnetic moment of the muon like we did for the 
electron, but taking into account the difference in masses.

{
\setbox0=\vbox{\hsize9.truecm{\epsfxsize 7.4truecm\epsfbox{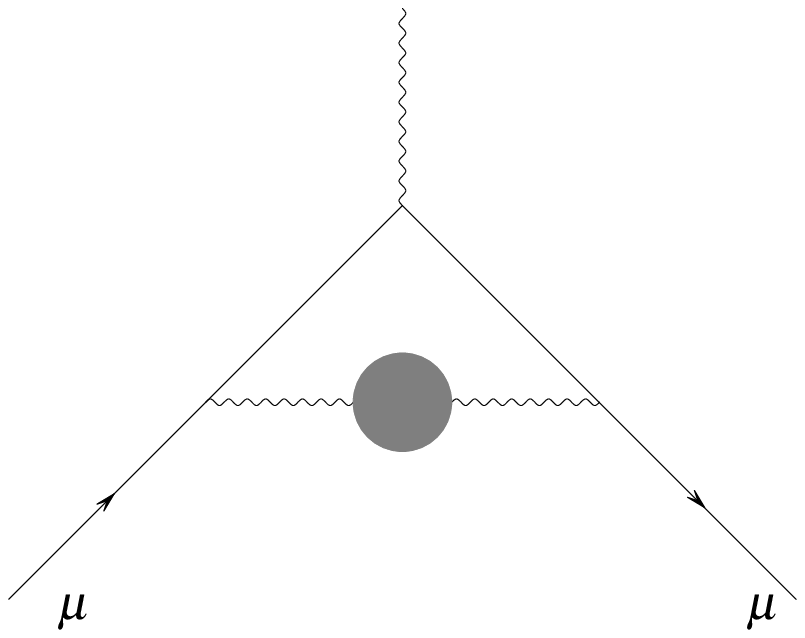}}} 
\setbox6=\vbox{\hsize 5.5truecm\captiontype\figuresc{Figure 5. }{The order $\alpha^2$
hadronic contributions to the muon magnetic moment. 
The blob represents an arbitrary hadronic state.\hb
\phantom{XX}}\hb
\vskip.1cm} 
\medskip
\line{\box0\hfil\box6}
\medskip
}

In 2001 a very precise measurement of the muon anomaly, $a(\mu)$, was 
carried over in Brookhaven.\ref{23a} 
It improved the previous CERN precision\ref{23b} by a factor of about six, 
and (according  to hasty 
voices) showed signs of 
discrepancy with experiment. 
This discrepancy was due to two factors. First, 
a somewhat overoptimistic and incomplete evaluation of the 
hadronic part of the photon vacuum polarization (h.v.p.) contribution, (\fig~5); and secondly, 
a sign mistake in one of the $O(\alpha^3)$ 
pieces, the so-called hadronic light by light diagram, 
\fig~6.

The various theoretical contributions may be 
grouped into three sets: 
purely QED, weak and hadronic. 
The first and second are\ref{1}
$$\eqalign{10^{11}\times a(\hbox{QED})=116\,584\,705&\pm1.8\cr
10^{11}\times a(\hbox{Weak})=151&\pm4.\cr
}\equn{(5.1)}$$

\bigskip
{
\setbox0=\vbox{\hsize9.truecm{\epsfxsize 7.truecm\epsfbox{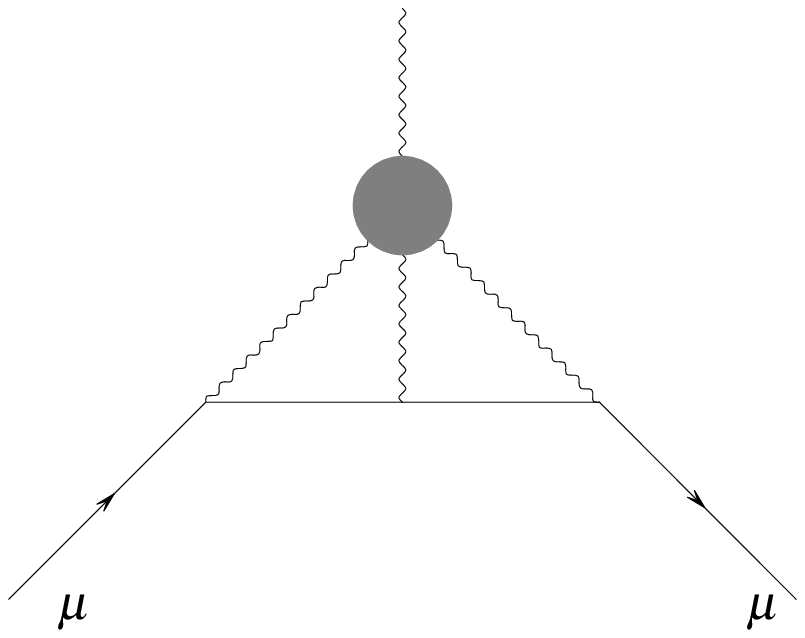}}} 
\setbox6=\vbox{\hsize 6.truecm\captiontype\figuresc{Figure 6. }{A 
typical diagram for the
 hadronic light by light contribution to 
  $a_\mu$.}\hb
\vskip.1cm} 
\medskip
\line{\box0\hfil\box 6}
\medskip
}
\goodbreak

The hadronic contributions may in turn be split  into four 
pieces: the $O(\alpha^2)$ hadronic photon vacuum polarization (h.v.p.)  
piece (\fig~5), and the same, to order $O(\alpha^3)$; we take both from 
ref.~24, which gives a much improved calculation with the 
help of very recent experimental data. We next have  
the light-by-light contribution (\fig~6), with 
the sign mistake 
redressed;\ref{25} and the rest, evaluated by Krause.\ref{26}
Thus, one has
$$\eqalign{10^{11}\times a^{(2+3)}(\hbox{h.v.p.})=&
7\,002\pm66;\cr
10^{11}\times a(\hbox{Hadronic light by light})= &92\pm20;\cr
10^{11}\times a(\hbox{Rest})=&-101\pm6.
\cr}
\equn{(5.2)}$$    
Once these added,  the theoretical value is\ref{24} 
$$a_\mu= (116 591 849\pm 72)\times 10^{-11}\quad\hbox{[Theory, standard model]},
\equn{(5.3)}$$
perfectly compatible (at $1.1\;\sigma$) with 
the experimental figure,
$$a_\mu= (116 592 030\pm150)\times 10^{-11}\quad\hbox{[Experiment]}.
\equn{(5.4)}$$
We may write the comparison, perhaps more transparently, as
$$a_\mu({\rm Exp.})-a_\mu({\rm Theory})=(1.81\pm1.66)\times 10^{-9}.
$$

This thus checks all three sectors of the
 the standard model (electromagnetic, weak and strong interactions are necessary to 
get the agreement) to nine digits. 
This is also the degree of accuracy of the verification of electron-muon universality.

\booksubsection{5.2.The electron dipole moment}
The interaction of a fermion, say the electron, with an 
electromagnetic field
through an electric and a magnetic 
moments may be described with the phenomenological interactions
$$-\tfrac{1}{2}\mu_e\bar{\psi}_e \sigma_{\mu\nu}F^{\mu\nu}\psi, \quad
\tfrac{1}{2}\ii d_e\bar{\psi}_e \sigma_{\mu\nu}F^{\mu\nu}\gamma_5\psi.
\equn{(5.3)}$$
We have already commented on the implications of the observed and calculated 
values of $\mu_e$, $\mu_\mu$. What about the dipole moment? 
Let us write $d_e=(e/m)\delta_e$. The present experimental bound\ref{27} on 
$d_e$, $2\times10^{-27}e\cdot{\rm cm}$, 
implies 
$$|\delta_e|\leq 5\times 10^{-16}.
\equn{(5.4)}$$ 
In the standard model, $\delta_e$ is zero, except for corrections of 
very high order, 
much smaller than the experimental bound; but this is not the case in other 
theories: therefore, 
the smallness of $d_e$ is a stringent test on 
structures beyond the standard model.
 Specifically, in the minimal suspersymmetric (SUSY) extension of the 
standard model the bound (5.4) implies a lower bound 
of the order of 1 \tev\ for the 
SUSY partners, if the SUSY CP-violating phases and mixing angles are of the 
same order of magnitude as the 
non-SUSY ones.

\vfill\eject
\booksection{Acknowledgements}
I am grateful to  P.~Langacker and 
M.~Gr\"unwald for information on the more recent value and estimate for 
 the masses of $t$ quark and 
Higgs particle. Thanks are also due to the organizers of 
the XXX International Meeting for the opportunity to talk here.

\booksection{References}
{

\item{1.-}{The values of $\alpha$ and the calculations of $a({\rm e})$, 
and (some of) those of $a_\mu$,
 with references, may be found in the review of 
\prajnyp{V.~W.~Hughes and T.~Kinoshita}{Rev. Mod.
Phys.}{71}{1999}{S133}.}
\item{2.-}{\/{\sc J. F. de Troc\'oniz and F. J. Yndur\'ain},  
FTUAM  01-15, 2001 (hep-ph/0107318), in press in Phys. Rev. D. 
A calculation with only slightly 
less good precision is that of \prajnyp{A.~D. Martin, 
J.~Outhwaite and M.~G.~Ryskin}{Eur. Phys. J.}{C19}{2001}{681}. 
Its central value is comprised within the error bars of the Troc\'oniz--Yndur\'ain 
result.}
\item{3.-}{\/PDG: \prajnyp{D. E. Groom et al.}{Eur. Phys. J.}{C15}{2000}{1}.}
\item{4.-}{For a review of this, see the text
  {\sc T.~Kinoshita et al.,} {\sl Quantum Electrodynamics}, World
Scientific, 1990.}
\item{5.-}{(a) $\tau$ decay: See 
{\sc A.~Pich,} {\sl Tau Physics}, In {\sl Heavy Flavours, II} (Buras and Lidner, eds.), 
World Scientific, Singapore,  (1997)
and work quoted there. (b) $Z$ decay: for a review, see 
\prajnyp{K.~G.~Chetyrkin,  J.~H.~Kuhn,, and  A.~Kwiatowski}
{Phys. Rep.}{C277}{1996}{189}. (c) $e^+e^-$ annihilations. 
The NNLO corrections 
calculated by
\prajnyp{S.~G.~Gorishny, A.~L.~Kataev and S.~A.~Larin}{Phys. Lett.}{B259}{1991}{144} 
and \prajnyp{L.~R.~Sugurladze, and M.~A.~Samuel}{Phys. Rev. Lett.}{66}{1991}{560}, 
where references to lower order ones may be found.}
\item{6.-}{The four loop coefficients for 
$\alpha_s$ were calculated by \prajnyp{T. van Ritbergen, J. A. M. Vermaseren and
 S. A. Larin}{Phys. Lett.}{B400}{1997}{379}. 
The results for the Bjorken and Gross--Llewellyn~Smith sum rules 
are taken from the review of \prajnyp{S. Bethke}{J. Phys.}{G26}{2000}{R27}.}
\item{7.-}{Four moments: \prajnyp{J.~Santiago and F. J. Yndur\'ain}{Nucl. Phys.}{B563}{1999}{45}. 
Six moments: \prajnyp{J.~Santiago and F. J. Yndur\'ain}{Nucl. Phys.}{B611}{2001}{447}. 
An evaluation for neutrino scattering was also given by 
{\sc A.~L.~Kataev, G.~Parente and A.~V.~Sidorov}, CERN TH-2001-058 
(hep-ph/0106221).
The calculations of the anomalous dimensions and 
coefficient functions at NNLO with respectively four and six moments  
used here come from \prajnyp{S. A. Larin et al.}{Nucl. Phys.}{B427}{1994}{41} and 
{\bf B492} (1997), 338; and 
\prajnyp{A. Retey and J. A. M. Vermaseren}{Nucl. Phys.}{B604}{2001}{281},
 where references to earlier work may be found. 
The perspectives for a full calculation 
for the anomalous dimensions are discussed by {\sc S.~Moch}, Proc. 
HEP Europhysics Conf., Budapest, 2001.}
\item{8.-}{\/{\sc D. Strom}, ``Electroweak measurements on the $Z$ resonance", 
Talk presented at the 5th Int. Symposium on Radiative Corrections, RadCor2000, 
Carmel, Ca., September 2000.  For discussion of the
Higgs, see also {\sc E.~Tournefier}, Talk at the Int. Seminar ``QUARKS '98",  
Suzdal (hep-ex/9810042).}
\item{9.-}{\prajnyp{M. Veltman}{Nucl. Phys.}{B123}{1977}{89}.}
\item{10.-}{\prajnyp{W. Fischler}{Nucl. Phys.}{B129}{1977}{157}; 
\prajnyp{A.~Billoire}{ Phys. Lett.}{B92}{1980}{343}.}
\item{11.-}{\prajnyp{S. Titard and F. J. Yndur\'ain}{Phys. Rev.}{D49}{1994}{6007}.}
\item{12.-}{\prajnyp{M. Peter}{Phys. Rev. Lett.}{78}{1997}{602}; 
checked and corrected by 
\prajnyp{Y.~Schr\"oder}{Phys. Lett.}{B447}{1999}{321}.}
\item{13.-}{\prajnyp{A. Pineda and F.~J.~Yndur\'ain}{Phys. Rev.}{D58}{1998}{3003} 
and \prajnyp{A. Pineda and F.~J.~Yndur\'ain}{Phys. Rev.}{D61}{2000}{077505}}
\item{14.-}{\prajnyp{M. B. Voloshin}{Nucl. Phys.}{B154}{1979}{365}  and 
{\sl Sov. J. Nucl. Phys.} {\bf 36}, 143  (1982); 
\prajnyp{H.~Leutwyler}{Phys. Lett.}{B98}{1981}{447}.}
\item{15.-}{\prajnyp{N.~Brambilla et al.,}{Phys. Lett.}{B470}{1999}{215};}
\item{16.-}{(a) {\sc F.~J.~Yndur\'ain}, 
 Proc. 11 Int. Seminar on High Energy Physics ``Quarks-2000", 
Pushkin, Rusia (hep-ph/0002237); (b) {\sc A.~H.~Hoang}, 
Proc. 
RADCOR2000, Carmel, Calif., 2000 (hep-ph/0102292).}
\item{17.-}{\prajnyp{R. Coquereau}{Phys. Rev.}{D23}{1981}{1365}, 
\prajnyp{R. Tarrach}{Nucl. Phys.}{ B183}{1981}{384} to one loop;
\prajnyp{N. Gray et al., }{Z. Phys.}{ C48}{1990}{673} to two loops; 
\prajnyp{K.~Melnikov and T.~van~Ritbergen}{Phys. Letters}{B482}{2000}{99}, 
to three loops.}
\item{18.-}{\/{\sc  F.~J.~Yndur\'ain}, {\sl The Theory of Quark and Gluon Interactions}, 3rd. ed., 
Springer Verlag, 1999.}
\item{19.-}{{\sc V. A. Novikov et al., {\sl Sov. J. Nucl. Phys.}, {\bf 27} (1978) 2, 274.}}
\item{20.-}{Bounds: {\sc C. Becchi S. Narison, E. de Rafael and F. J. Yndur\'ain, \jzp{C8}{1981}{335.}}; 
\prajnyp{F.~J.~Yndur\'ain}{Nucl. Phys.}{B517}{1998}{324}. 
Ratios and estimates: {\sc D.~B.~Kaplan and A.~V.~Manohar \jprl{56}{1986}{2004}}; 
\prajnyp{K.~G.~Chetyrkin, D.~Pirjol and K.~Schilcher}{Phys. Lett.}{B404}{1997}{337};
\prajnyp{J.~Bijnens, J.~Prades and E.~de~Rafael}{Phys. Lett.}{B348}{1995}{226.}}
\item{21.-}{{\sc D. J. Broadhurst}, \jpl{101}{1981}{423;} 
{\sc K.~G.~Chetyrkin, D.~Pirjol and K.~Schilcher}, \jpl{B404}{1997}{337};
{\sc S.~C.~Generalis}, {\sl J. Phys.}, {\bf G16} (1990) 787; {\sc L.~R.~Sugurladze, 
and F.~V.~Tkachov}, \jnp{B331}{1990}{35};
 {\sc K.~G.~Chetyrkin, S.~G.~Groshny and F.~V.~Tkachov}, \jpl{119}{1982}{407}; 
{\sc S.~G.~Groshny, A.~L.~Kataev, S.~A.~Larin and L.~R.~Sugurladze}, 
\jpr{D43}{1991}{1633}; {\sc P. Pascual and E. de Rafael}, \jzp{12}{1982}{127}; 
{\sc J.~A.~M.~Vermaseren et al.}, \jpl{B405}{1997}{327}; {\sc K.~G.~Chetyrkin}, 
\jpl{B404}{1997}{161}.}
\item{22.-}{See \prajnyp{S.~Chen}{Nucl. Phys. B (Proc. Suppl.)}{64}{1998}{265} 
for a recent review of the experimental situation.}
\item{23.-}{BNL: \prajnyp{H.N. Brown et al.}{Phys. Rev. Letters}{86}{2001}{2227}. 
CERN: \prajnyp{J. Bailey et al.}{Nucl. Phys.}{B150}{1979}{1}.}
\item{24.-}{\/{\sc J. F. de Troc\'oniz and F. J. Yndur\'ain}, FTUAM 01-08 
(hep-ph/0106025), in press in Phys. Rev.~D.}
\item{25.-}{\/{\sc M.~Knecht and A.~Nyffeler} (hep-ph/0111058), 
confirmed in {\sc M.~Hayakawa and T.~Kinoshita} (hep-ph/0112102).}
\item{26.-}{\prajnyp{B. Krause}{Phys. Letters}{B390}{1997}{392}.}
\item{27.-}{\prajnyp{E. D. Commins et al.}{Phys. Rev.}{A50}{1994}{2960}.}
\item{}{}
}
\bye